\documentclass[twocolumn,aps,pra,showpacs]{revtex4}
\usepackage{epsfig}
\begin{document}
\title{Conditional implementation of asymmetrical universal quantum cloning machine}
\author{Radim Filip}
\affiliation{Department of Optics, Palack\' y University,\\
17. listopadu 50,  772~07 Olomouc, \\ Czech Republic}
\date{\today}
\begin{abstract}
We propose two feasible experimental implementations of an optimal 
asymmetric $1\rightarrow 2$ quantum cloning of a polarization state of photon. 
Both implementations are based on a partial and optimal reverse of recent conditional 
symmetrical quantum cloning experiments. The reversion procedure is performed only by a local measurement 
of one from the clones and ancilla followed by a local operation on the other clone.
The local measurement consists only of a single unbalanced beam splitter followed in one output by a single photon detector 
and the asymmetry of fidelities in the cloning is controlled by a reflectivity of the beam splitter. 
\end{abstract}
\maketitle


Copying the quantum states is apparently dissimilar to 
classical information processing since it is impossible to precisely 
duplicate an unknown quantum state as a consequence of a 
linearity of quantum mechanics \cite{nocloning}. 
To clone an unknown quantum 
state at least approximately, universal quantum cloning machines (UQCM) were developed \cite{UQCM}.
The UQCM is a device that universally and optimally produces a copy $\rho_{S'}$ from an 
unknown quantum state $|\Psi\rangle_{S}$ of the original. 
Specifically, an optimal symmetrical $1\rightarrow 2$ UQCM (SUQCM) for qubits 
creates a copy $\rho_{S'}$ with a maximal state-independent fidelity $F_{S'}=_{S}\langle\Psi|\rho_{S'}|\Psi\rangle_{S}=5/6$. 
Simultaneously, a pure state of original changes to mixed state $\rho_{S}$ exhibiting maximally the same fidelity $F_{S}=
_{S}\langle\Psi|\rho_{S}|\Psi\rangle_{S}=5/6$ as the clone. 
To optimally control the fidelities $F_{S}$ and $F_{S'}$, a concept 
of asymmetrical UQCM (AUQCM) has been theoretically developed \cite{Niu98,Buzek98,Cerf00}. 
The optimal $1\rightarrow 2$ AUQCM produces the copies having state-independent fidelities controlled by a 
parameter $R$ in such a way that for a given fidelity of the copy $F_{S'}(R)$, the fidelity $F_{S}(R)$ of the 
original is maximal. More specifically, assuming a qubit in an unknown state $|\Psi\rangle$ then 
the original $S$ and clone $S'$ leaving $1\rightarrow 2$ AUQCM can be represented by the following density matrices 
\cite{Niu98,Buzek98,Cerf00}
\begin{eqnarray}
\rho_{S,S'}=F_{S,S'}|\Psi\rangle\langle\Psi|+(1-F_{S,S'})|\Psi_{\bot}\rangle\langle\Psi_{\bot}|
\end{eqnarray}
where the fidelities 
$5/6\leq F_{S}\leq 1$ and $1/2\leq F_{S'}\leq 5/6$ satisfy the cloning relation 
\begin{equation}\label{cond}
(1-F_{S})(1-F_{S'})\geq (1/2-(1-F_{S})-(1-F_{S'}))^{2},
\end{equation}
where the equality corresponds to an {\em optimal} AUQCM, in the sense that for a a larger $F_{S}$ cannot be 
obtained for given $F_{S'}$. The Eq. (\ref{cond}) is the tightest no-cloning bound for the fidelities 
of the $1\rightarrow 2$ cloner which copies an unknown qubit state to the another with an isotropic noise.  
The recent experimental effort to build different quantum cloners is mainly stimulated 
by their use as individual attacks in quantum communication and cryptography \cite{attack}. 
More information about this practical application of the asymmetrical universal cloning as optimal attack 
for a cryptographic protocol can be found in Ref.~\cite{asymmattack}. 

To build quantum cloners, quantum networks using CNOT gates for both optimal SUQCM and AUQCM were proposed \cite{netclon}.
However, a strength of the state-of-the-art nonlinear interaction at a single photon 
level is unfortunately too weak to produce a deterministic and 
efficient CNOT operation only by a direct interaction
between photons. For this reason, the deterministic SUQCM and AUQCM still have not been experimentally 
implemented in quantum optics. Netherless, stimulated or
spontaneous parametric down-conversions were used to realize a conditional implementation 
of the SUQCM for a polarization state of photon \cite{Lamas-Linares02,DeMartini02,Ricci03}. 
However, to the best of our knowledge, no feasible experimental setup for an 
optimal $1\rightarrow 2$ AUQCM has been presented yet. On the other hand, an experimental realization of 
optimal asymmetrical cloning machine 
has already been proposed for coherent states \cite{CVasymm}. 

In this paper, we propose optimal $1\rightarrow 2$ AUQCM which is a simple and experimentally feasible extension of 
the recent experiments on the conditional symmetrical cloning of the polarization state of a photon. 
Our method is based on a partial optimal reverse of SUQCM by a specific controllable 
joint measurement on one of the copies and an auxiliary photon leaving the cloning process. 
By this partial reverse the quantum information between the disturbed original and copy can be redistributed posteriori
only using the local operations and classical communication. It can be experimentally accomplished adding only 
a single unbalanced beam splitter followed by a single-photon detector 
in the recent cloning experiments \cite{Lamas-Linares02,DeMartini02,Ricci03}. 

\begin{figure}
\centerline{\psfig{width=8.0cm,angle=0,file=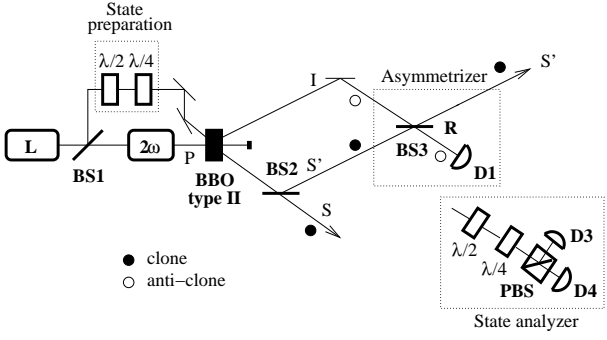}}
\caption{Setup for conditional AUQCM based on stimulated parametric down-conversion: L -- laser, BS1-BS3 beam splitters, PBS -- polarization beam splitter, 2$\omega$ -- frequency doubler, BBB -- nonlinear 
BBO type II crystal, $\lambda/2,\lambda/4$ -- wave plates, D1 -- single photon detector.}
\end{figure}

An experimental realization of the conditional 1$\rightarrow$2 SUQCM \cite{DeMartini02,Lamas-Linares02} 
was based on a nonlinear parametric down-conversion stimulated in a signal beam by 
a single photon prepared in an unknown polarization state 
$|\Psi\rangle_{S}=a|V\rangle_{S}+b|H\rangle_{S}$, where $V$ and $H$ 
denote the vertical and horizontal polarizations. 
The experimental arrangement is depicted in Fig.~1. The input single 
photon extracted from a laser pulse is prepared
in the state $|\Psi\rangle_{S}$ in a preparation device using $\lambda/2$ and $\lambda/4$ wave plates. 
A more intensive part of the laser pulse is frequency doubled and used to
pump a BBO non-linear crystal. An action of a non-degenerate 
type II parametric down-conversion process in the crystal can be described by the Hamiltonian
$H_{I}=i\hbar\chi(a^{\dag}_{H}b^{\dag}_{V}-a^{\dag}_{V}b^{\dag}_{H})+\mbox{h.c.}$, 
where $\chi$ is proportional to 
a nonlinear susceptibility of the crystal, and h.c. denotes the hermitian conjugation. 
Here the annihilation operators $a,b$ are assumed to be acting on the selected signal mode $S$ 
and idler mode $I$, respectively. A short-time 
approximation of the evolution operator 
$U=\exp(-iH_{I}t/\hbar)\approx 1-\frac{it}{\hbar}H_{I}$ is used. This approximation 
is correct for this kind of the experiments, since the gain $g=\chi t$ is usually very small $(|g|\ll 1)$. Within 
the short-time approximation, polarization basis states $|H\rangle_{S}\equiv |1,0\rangle_{S}$ and 
$|V\rangle_{S}\equiv |0,1\rangle_{S}$ of the input photon evolve according to the following rules
\begin{eqnarray}
U|1,0\rangle_{S}|0,0\rangle_{I}&\approx& |1,0\rangle_{S}|0,0\rangle_{I}+g(\sqrt{2}|2,0\rangle_{S}|0,1\rangle_{I}\nonumber\\
& &-|1,1\rangle_{S}|1,0\rangle_{I},\nonumber\\
U|0,1\rangle_{S}|0,0\rangle_{I}&\approx& |0,1\rangle_{S}|0,0\rangle_{I}-g(\sqrt{2}|0,2\rangle_{S}
|1,0\rangle_{I}\nonumber\\
& &-|1,1\rangle_{S}|0,1\rangle_{I}.
\end{eqnarray}
Here, the produced states $|2,0\rangle$ and $|0,2\rangle$ 
represent an effect of a stimulated emission which is used to prepare 
the clone, and the state $|1,1\rangle$ corresponds to an unavoidable 
effect of a spontaneous emission. After the amplification, 
a balanced polarization-insensitive beam splitter $BS2$ in the signal mode 
separates two photons in the states $|2,0\rangle$ or $|0,2\rangle$ to two distinguishable
spatial modes corresponding to the disturbed photon $S$ and clone $S'$. 
An action of the beam splitter $BS2$ on a pair of photons is as follows
\begin{eqnarray}\label{BS}
|1,1\rangle_{S}|0,0\rangle_{S'}&\rightarrow &\frac{1}{2}(|1,1\rangle_{S}|0,0\rangle_{S'}+
|0,0\rangle_{S}|1,1\rangle_{S'}+\nonumber\\
& &|1,0\rangle_{S}|0,1\rangle_{S'}+|0,1\rangle_{S}|1,0\rangle_{S'}),\nonumber\\
|2,0\rangle_{S}|0,0\rangle_{S'}& \rightarrow &\frac{1}{2}(|2,0\rangle_{S}|0,0\rangle_{S'}+|0,0\rangle_{S}|2,0\rangle_{S'})+
\nonumber\\
& &\frac{1}{\sqrt{2}}|1,0\rangle_{S}|1,0\rangle_{S'},\nonumber\\
|0,2\rangle_{S}|0,0\rangle_{S'} &\rightarrow &\frac{1}{2}(|0,2\rangle_{S}|0,0\rangle_{S'}+|0,0\rangle_{S}|0,2\rangle_{S'})+
\nonumber\\
& &\frac{1}{\sqrt{2}}|0,1\rangle_{S}|0,1\rangle_{S'},
\end{eqnarray}
where $S$ and $S'$ are the signal modes. In the next 
procedure only such cases when a single 
photon is present in the mode $S$ are considered. 
Returning to the previous notation $|1,0\rangle_{i}=|H\rangle_{i}$ and 
$|0,1\rangle_{i}=|V\rangle_{i}$, the SUQCM transformation 
\begin{eqnarray}\label{clontr}
|\Psi\rangle_{S}\rightarrow \sqrt{\frac{2}{3}}|\Psi\Psi\Psi^{\bot}\rangle_{SS'I}-\frac{1}{\sqrt{3}}
|\Psi_{+}\rangle_{SS'}|\Psi_{I}\rangle,
\end{eqnarray} 
where $|\Psi_{+}\rangle=\frac{1}{\sqrt{2}}
(|\Psi\Psi^{\bot}\rangle_{SS'}+|\Psi^{\bot}\Psi\rangle_{SS'})$ and $|\Psi^{\bot}\rangle=a^{*}
|H\rangle-b^{*}|V\rangle$ is the orthogonal state to $|\Psi\rangle$, is actually performed. This SUQCM is optimal and transforms 
an unknown state $|\Psi\rangle_{S}$ of the original to the disturbed one and a copy, with the fidelities $F_S=F_S'=5/6$. 
Both the output states of photons $S$ and $S'$ have to be measured using the state analyzer composed from the 
$\lambda/2$-wave plate and $\lambda/4$-wave plate, the polarization beam splitter $PBS$ 
and a pair of single photon detectors $D3,D4$, as depicted in Fig.~1. 
Note, in the cloning experiment \cite{Lamas-Linares02},  
the fidelities were approximately $F_{S},F_{S'}\approx 0.81$ which are really close to the optimal value of $5/6=0.833$.



\begin{figure}
\vspace{1cm}
\centerline{\psfig{width=8.0cm,angle=0,file=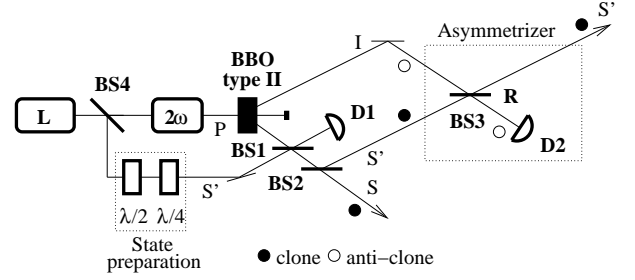}}
\caption{Setup of conditional AUQCM based on spontaneous parametric down-conversion: 
L -- laser, BS1-BS3 beam splitters, PBS -- polarization beam splitter, 2$\omega$ -- frequency doubler, BBB -- nonlinear 
BBO type II crystal, $\lambda/2,\lambda/4$ -- wave plates, D1-D2 -- single-photon detectors.}
\end{figure}

Recently, a different implementation of the $1\rightarrow 2$ SUQCM was experimentally performed \cite{Ricci03}. 
It is based on the following joint projection 
\begin{equation}\label{sproj}
\Pi_{S}=(1_{SS'}-|\Psi_{-}\rangle_{SS'}\langle\Psi_{-}|)\otimes 1_{I}
\end{equation}
of an unknown polarization state $|\Psi\rangle_{S'}=a|V\rangle+B|H\rangle$ of the input photon and  
the antisymmetric polarization Bell state 
$|\Psi_{-}\rangle_{SI}=\frac{1}{\sqrt{2}}(|VH\rangle-|HV\rangle)$ of two photons produced by the spontaneous parametric 
down-conversion from the same BBO nonlinear crystal as in the previous experiment. 
The corresponding experimental setup is depicted in Fig.~2. 
The initial state $|\Psi\rangle_{S'}=a|V\rangle_{S'}+b|H\rangle_{S'}$ is prepared by the same method.
The projection $\Pi_{S}$ to the symmetric subspace \cite{Werner98} on the state $|\Psi\rangle_{S'}|\Psi_{-}\rangle_{SI}$ 
can be accomplished by a sequence of two beam splitters $BS1$, $BS2$ and 
a single-photon detector $D1$. If two input photons in the same state 
$|0,1\rangle_{S}|0,1\rangle_{S'}$ or $|1,0\rangle_{S}|1,0\rangle_{S'})$ 
constructively interfere on the first balanced  
beam splitter $BS1$ and no photon is detected by the detector $D1$, then a twin of photons in the state $|0,2\rangle_{S}$ or 
$|2,0\rangle_{S}$ is produced in the mode $S$ with a success probability $1/2$.
The twins are further divided on the second balanced beam splitter 
$BS2$ to separate the photons to the different spatial modes and if we selected only such cases when 
exactly an single photon is in every mode $S,S',I$ the symmetric states
$|0,1\rangle_{S}|0,1\rangle_{S'}$ or $|1,0\rangle_{S}|1,0\rangle_{S'}$ with probability $1/4$ are obtained at a result.  
On the other hand, if two orthogonal basis states $|0,1\rangle_{S}|1,0\rangle_{S'},|1,0\rangle_{S}
|0,1\rangle_{S'}$ are mixed at the beam splitter $BS1$, they do not mutually interfere and in addition,
if no photon is registered on the detector $D1$, then the state $|1,1\rangle_{S}$ with the 
success probability $1/2$ is within the mode $S$.  
Thus after splitting the photons by $BS2$ to separate spatial modes, 
a symmetric state $\frac{1}{\sqrt{2}}(|1,0\rangle_{S}|0,1\rangle_{S'}+|0,1\rangle_{S}|1,0\rangle_{S'})$ is obtained with 
the total success probability $1/2$. Thus, with the probability $1/4$ the following 
transformation 
\begin{eqnarray}\label{symm}
|\Psi\Psi\rangle_{SS'}&\rightarrow & \sqrt{2}
|\Psi\Psi\rangle_{SS'},\nonumber\\
|\Psi\Psi_{\bot}\rangle_{SS'}&\rightarrow &\frac{1}{\sqrt{2}}(
|\Psi\Psi_{\bot}\rangle_{SS'}+|\Psi_{\bot}\rangle_{SS'}),
\end{eqnarray}
of the states of the photons $S,S'$ is in fact conditionally implemented. Assuming that a state of the idler photon $I$ 
is selected only if this procedure is successful, the total projection 
(\ref{sproj}) transforms the input state $|\Psi\rangle_{S'}|\Psi_{-}\rangle_{SI}$ to (\ref{clontr}). Thus 
the optimal SUQCM is conditionally accomplished however now a spontaneous emission of maximally entangled pairs is used rather 
than a stimulated emission in the previous experiment.  
In the experiment based on this idea \cite{Ricci03}, the fidelities of the clone and disturbed original are 
$F_{S},F_{S'}\approx 0.826$ which are even more closer to the theoretical value $5/6=0.833$ than it has been in the previous case.


An extension of both setups to achieve the optimal AUQCM can be presented. 
It is known that the symmetrical quantum cloning is LOCC reversible \cite{Bruss01}. 
If a projective measurement $\Pi_{-}=|\Psi_{-}\rangle\langle\Psi_{-}|$ on one clone and ancilla is performed, 
the other clone returns back to the initial state $|\Psi\rangle$. 
Thus we can guess that if an appropriate projection in a form 
$\alpha 1+\beta \Pi_{-}$ is applied on the one clone and the ancilla, an intermediate case
corresponding to the optimal asymmetrical cloning machine could be obtained.  

Now we show that this projection can be conditionally implemented if one mixes a pair of photons in the idler $I$ 
and signal $S'$ modes on unbalanced beam splitter $BS3$ having a variable reflectivity $0\leq R\leq 1/2$ and 
select only the cases when both photons leaving the beam splitter are separated. Then  
the beam splitter can be simply described by transformation 
\begin{eqnarray}\label{unbaltr}
|VV\rangle_{S'I} &\rightarrow &(T-R)|VV\rangle_{S'I},\nonumber\\
|HH\rangle_{S'I} &\rightarrow & (T-R)|HH\rangle_{S'I},\nonumber\\
|HV\rangle_{S'I}&\rightarrow & T|HV\rangle_{S'I}-R|VH\rangle_{S'I},\nonumber\\
|VH\rangle_{S'I}&\rightarrow & T|VH\rangle_{S'I}-R|HV\rangle_{S'I},
\end{eqnarray}
where $T+R=1$. It can be simply proved that this transformation can be expressed in a covariant 
way 
\begin{eqnarray}
|\Psi\Psi\rangle_{S'I} &\rightarrow & (T-R)|\Psi\Psi\rangle_{S'I},\nonumber\\
|\Psi\Psi_{\bot}\rangle_{S'I}&\rightarrow & T|\Psi\Psi_{\bot}\rangle_{S'I}-R|\Psi_{\bot}\Psi
\rangle_{S'I}.
\end{eqnarray}
It is an asymmetrical projection controlled by the parameter $R$, 
in a contrast to the symmetrizing projection (\ref{symm}). If an output state is selected only when there is 
an single photon in each mode $S,S',I$, we obtain the following transformation 
for the polarization states of the photon
\begin{eqnarray}\label{trans1}
|H\rangle &\rightarrow &\frac{1}{\sqrt{N(R)}}
((2-R)|HHV\rangle_{SS'I}-\nonumber\\
& &(1+R)|HVH\rangle_{SS'I}-
(1-2R)|VHH\rangle_{SS'I}),\nonumber\\
|V\rangle &\rightarrow &\frac{1}{\sqrt{N(R)}}((2-R)|VVH\rangle_{SS'I}-\nonumber\\
& &(1+R)|VHV\rangle_{SS'I}-
(1-2R)|HVV\rangle_{SS'I})\nonumber\\
\end{eqnarray}
where $N(R)=6(1-R(1-R))$. In a real experiment, 
a detection of the photon in the mode $I$ can be performed destructively by a single-photon detector $D2$ whereas 
the signal photons from total cloning operation are detected in the state analyzers. 
It can be simply proved that the transformation (\ref{trans1}) 
is covariant and it can be written in a form 
\begin{eqnarray}\label{trans1}
|\Psi\rangle &\rightarrow &\frac{1}{\sqrt{N(R)}}
((2-R)|\Psi\Psi\Psi_{\bot}\rangle_{SS'I}-\nonumber\\
& &(1+R)|\Psi\Psi_{\bot}\Psi\rangle_{SS'I}-
(1-2R)|\Psi_{\bot}\Psi\Psi\rangle_{SS'I})\nonumber\\
\end{eqnarray}
which corresponds to the following projection 
\begin{equation}
\Pi_{A}(R)=\left((1-2R)1_{S'}\otimes 1_{I}+2R|\Psi_{-}\rangle_{S'I}\langle\Psi_{-}|\right)\otimes 1_{S},
\end{equation}
on the state (\ref{clontr}) produced by the SUQCM. 
An interpretation of this projective measurement is straightforward: the asymmetrical cloning is obtained 
as a partial optimal reverse of the symmetrical one. For $R=0$ the SUQCM is obtained and for $R=1/2$ the SQUCM is reversed and 
an initial state of the original is precisely restored. 
An optimal total reverse of the state after the symmetrical cloning was previously 
theoretically already analyzed \cite{Bruss01}. 
After the total reversion, any input state 
is deterministically revealed by the complete Bell-state measurement on the clone and ancilla.
Thus this obtained result also represents a solution of the problem of a partial but still optimal reversion of 
the symmetrical cloning. Further, this reversion is also obtained only using the 
local operations on the clones and classical communication 
between them. It enables the redistribution of the quantum information encoded in symmetric clones at a 
distance without an additional quantum channel.  
 

\begin{figure}
\vspace{1cm}
\centerline{\psfig{width=8.0cm,angle=0,file=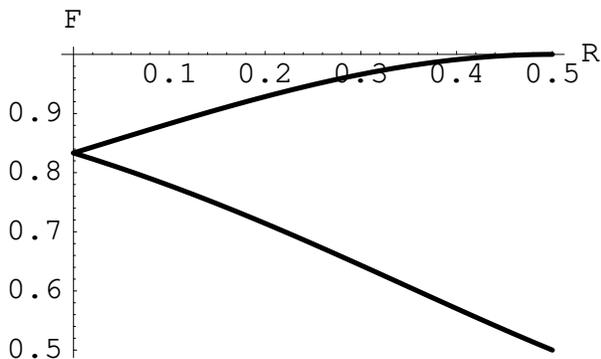}}
\caption{Fidelity of original and clone after AUCQM in dependence on beam splitter $BS3$ reflectivity. 
The upper curve corresponds to $F_{S}$ and the lower curve to $F_{S'}$}
\end{figure}

Since the photon in the idler mode is detected in such a way 
that no information about its polarization is acquired, we trace over the idler mode and obtain 
the final output states of the modes $S$ and $S'$ 
\begin{eqnarray}
\rho_{SS'}&=&\frac{1}{N(R)}
\left[(2-R)^{2}|\Psi\Psi\rangle_{SS'}\langle\Psi\Psi|+\right.\nonumber\\
& &\left.\left((1+R)|\Psi\Psi_{\bot}\rangle_{SS'}+
(1-2R)|\Psi_{\bot}\Psi\rangle_{SS'}\right)\times\right.\nonumber\\
& &\left.\left((1+R)\langle\Psi\Psi_{\bot}|_{SS'}+
(1-2R)\langle\Psi_{\bot}\Psi|_{SS'}\right)\right].\nonumber\\
\end{eqnarray}
This state carries both the disturbed original and clone 
\begin{eqnarray}\label{final}
\rho_{S}&=&\frac{1}{N(R)}(((2-R)^{2}+(1+R)^{2})|\Psi\rangle\langle\Psi|+\nonumber\\
& &(1-2R)^{2}|\Psi_{\bot}\rangle\langle\Psi_{\bot}|),\nonumber\\
\rho_{S'}&=&\frac{1}{N(R)}(((2-R)^{2}+(1-2R)^{2})|\Psi\rangle\langle\Psi|+\nonumber\\
& &(1+R)^{2}
|\Psi_{\bot}\rangle\langle\Psi_{\bot}|)
\end{eqnarray}
and is conditioned by a simultaneous detection of a single photon in each of the output modes $S,S',I$.
The marginal states of the disturbed original and clone in the selected sub-ensemble have the 
following fidelities 
\begin{eqnarray}\label{fid}
F_{S}&=&\frac{(2-R)^{2}+(1+R)^{2}}{6(1-R(1-R))},\nonumber\\
F_{S'}&=&\frac{(2-R)^{2}+(1-2R)^{2}}{6(1-R(1-R))},
\end{eqnarray}
with an initial state $|\Psi\rangle$ which vary with increasing $R\in \langle 0,1/2\rangle$ 
from a perfect SUQCM $(R=0)$ to the trivial non-cloning case $(R=1/2)$, as depicted in Fig.~3. 
The output states of the original and clone can be measured by the state analyzers analogically as it was discussed for 
the experiments with the SUQCMs. Inserting the fidelities (\ref{fid}) to the cloning inequality (\ref{cond}) which restricts 
all the possible AUQCM, the equality is obtained in Eq.~(\ref{cond}) as can be straightforwardly proved.
 
In this paper an extension of the recent conditional 
cloning experiment for a polarization state of photon toward the optimal asymmetrical $1\rightarrow 2$ quantum cloning machine 
is proposed. 
Our method is based on a conditional partial optimal reverse of the SUQCM controlled by an experimental parameter $R$. 
We have applied this method in the recent symmetrical cloning 
experiments \cite{Lamas-Linares02,DeMartini02,Ricci03} to obtain the optimal asymmetrical cloning.   
In summary, the AUQCM can be described as the projection $\Pi(R)=\Pi_{A}(R)\Pi_{S}$, given explicitly by
\begin{equation}
\Pi(R)=\left((2-R)1_{S'}\otimes 1_{S}-2(1-2R)|\Psi_{-}\rangle_{S'S}\langle\Psi_{-}|\right)\otimes 1_{I},
\end{equation} 
on the state $|\Psi\rangle_{S'}\otimes |\Psi_{-}\rangle_{SI}$ composed from an initial unknown 
state and the antisymmetric Bell state produced from the spontaneous parametric down-conversion.
Since the fidelities obtained in these experiments  
are very close to the optimal value $5/6$ and the proposed modification 
is rather simple, this asymmetrical cloning machine is feasible and it could be 
straightforwardly realized.

\medskip
\noindent {\bf Acknowledgments}
The work was supported by the Post-Doc project 202/03/D239 of Grant agency of Czech Republic and
projects LN00A015 and CEZ: J14/98 of the Ministry of Education of the Czech Republic. I would like to
thank Jarom\' ir Fiur\' a\v sek and Petr Marek for the stimulating and fruitful discussions.


\end{document}